**Author's Original Manuscript**

# Towards a Framework for Enterprise Architecture in Mobile Government: A Case Study

[Do not insert author name or contact details now - only to be completed after acceptance]

**Abstract:** Mobile government (m-government) represents a distinct paradigm shift from electronic government (e-government), offering a new avenue for governments worldwide to deliver services and applications to their customers. The m-government model deviates from e-government in terms of information technology (IT) infrastructure, security, and application management and implementation. Enterprise architecture (EA) has been developed and utilized globally to enhance efficiency and information and communication technology (ICT) utilization in the public sector through e-government. However, the application of EA within the context of m-government, particularly in developing countries, has largely been overlooked by scholars. This study aims to address this gap. This study seeks to develop an EA specifically tailored for m-government in a developmental context. Our contribution to the literature is the illustration of a proposed EA framework for m-government. The practical implementation of this study is to identify critical considerations when designing and adopting m-government to avoid redundant investments during the integration of infrastructure and applications from e-government to m-government.

**Biographical Notes:**

**Author 1**.

**Author 2**.

**Author 3**.

**Author 4**.

## 1 Introduction

Interoperability is a key objective in the development of information systems (IS) within governmental structures (Otjacques et al., 2007). It facilitates the alignment of business processes, responsibilities, and expectations to achieve mutually beneficial goals, with a focus on user needs (EC, 2024). Thus, governments worldwide are exploring strategies to design IS that enhance interoperability. One such strategy is Enterprise Architecture (EA), which refers to methodologies aimed at improving the alignment between an organization's business and its IS (Dang and Pekkola, 2017). In the public sector, EA is often associated with electronic government (e-government), also known as e-government EA.

In the era of digital technologies (e.g., 5G, IoT, blockchain, social media), e-government is evolving into digital government (Dang and Vartiainen 2022). This trend is driven by the proliferation of smart devices and the preference of customers to use services offered by the government on these devices, as opposed to traditional computer-based services in e-government. This shift presents both opportunities and challenges for governments. On one hand, providing services via smart devices can enhance communication between governments and citizens (Al-Hubaishi, 2010; Abdalla, 2013), thereby transforming traditional e-government into mobile government (m-government) (Kanaan et al., 2019). In other words, m-government change the way to provide services public (Sareen et al., 2013). On the other hand, the rapid evolution of technologies and their configurations and infrastructure introduces new interoperability challenges. In this context, EA for m-government is considered a viable approach to address these challenges.

Numerous scholars and countries have conducted studies and developed EA for mobile government (Ala'a and Roesnita, 2020). Discussions have encompassed a variety of topics, including success factors (Ishengoma et al., 2019), components of frameworks (Alharbi et al., 2020), and challenges (Sareen et al., 2013), among others. However, there is a gap in the literature regarding the development of a comprehensive EA framework for mobile government that spans the entire lifecycle from initiation to design, implementation, and management, especially in the developing countries context. Our research aims to address this gap by focusing on the development of an EA framework for mobile government.

We initially proposed a framework, drawing upon a literature review and the consideration of widely recognized EA frameworks such as TOGAF and FEAF. Subsequently, we employed the Theory of Planned Behavior (TPB) (Ajzen, 1991) to predict a user's intention to utilize the proposed frameworks. We then examined the compatibility of our proposed framework with existing ones and its applicability within the study context. Vietnam was selected as the study context for three primary reasons. First, the number of mobile phone subscribers in Vietnam exceeds its total population. Moreover, Vietnam has already established an EA for e-government and is in the process of transitioning from e-government to digital government, which includes m-government. The government is also promoting the provision of services via smart devices and encouraging the use of digital technologies to enhance service quality (Government Decision, 2020). This environment allows us to address the research gaps. Second, literature indicates that not all components of e-government are applicable in m-government, and m-government possesses unique features that distinguish it from e-government (Ghazali and Razali, 2014). Thus, the Vietnamese government needs to revise its EA for e-government to develop a new EA for m-government. To the best of our knowledge, the Vietnamese government has not yet developed an EA for m-government, and there has been no research conducted on m-government in Vietnam. This presents potential practical implications for practitioners in Vietnam. Third, we had the opportunity to access not only secondary data but also conduct interviews and research activities in Vietnam.

We make a contribution to the literature by presenting an EA for m-government, specifically tailored to the context of developing countries. Our proposed framework is comprehensive, encompassing three primary phases: the initiation phase, the design and implementation phase, and the management phase. Each phase is further subdivided into its constituent layers or components. This framework serves as a valuable reference for practitioners, particularly when establishing EA for m-government within state agencies.

This article is structured as follows: Following this introduction, Section 2 provides a review of the relevant literature. Section 3 outlines the research methods employed, while Section 4 presents the findings of the research. In Section 5, we discuss the comparability and applicability of the proposed framework. Finally, Section 6 concludes the article and discusses potential limitations.

## 2. Literature review

### *2.1 E-government and enterprise architecture*

There is no globally accepted definition of e-government (Halchin, 2004). As it is often understood either as a utilization of ICT, an adoption of Internet, a methodology, and perhaps even all of these simultaneously for delivering government services to its constituents (Grönlund and Horan, 2005). In this paper, we refer e-government as "utilizing the Internet and the World-Wide-Web for delivering government information and services to citizens" (United Nations, 2002). Similar to e-government, EA is understood to be different from different scholars, governments and practitioners. EA is "an approach to improve the alignment between the organization's business and their information technologies" (Dang and Pekkola, 2017).

EA's studies in the public sector can be categorized into three main streams: EA development, EA adoption, and EA benefits (Table 1) (Dang and Pekkola, 2017). The first stream focuses on development of enterprise architecture frameworks and its related issues in the public sector, such as EA maturity models, EA evaluation, EA assessment, as well as frameworks for interoperability, integration. The scopes of this stream can be general, international context, central government context, local government context or line of businesses (e.g., healthcare, lands management, or social services). The second research stream focuses on EA adoption. This steam dive into how public agencies adopt or use EA, as well as challenges and solutions to overcome EA adoption, such as EA adoption problems, root causes of EA adoptions, and institutionalization process of EA adoption. The third trend focuses on the benefits of EA to discuss EA benefit or EA benefit realization or organizational benefits of EA.

**Table 1.** EA research in the public sector.

| Research stream | Main focus | Selected References |
|---|---|---|
| EA development | EA concepts, EA frameworks | (Halchin, 2004; Zachman, 1987) |
| EA adoption | How EA has been used in the public sectors | (Dang 2017; Dang, 2019) |
| EA benefits | How EA benefits and realized in the public sector | (Foorthuis et al. 2016; Tamm et al. 2011) |

These streams of EA research in the public sector focus on EA for e-government, but not, for example, m-government. Also, most of EA studied are in the developed context, where there is lacks study in the developing countries. As a result, there is a need for more research on EA development in a development context (Dang and Pekkola, 2017). This is the aim of the study, with particular emphasis on EA for m-government.

### *2.2. M-government and its challenges*

M-Government is conceptualized as a strategy and its corresponding implementation by the government, utilizing mobile devices to provide information, deliver services, engage citizens, and enhance efficiency (Lee et al., 2006). It is indicated that m-government can confer additional benefits to e-government in a multitude of ways (Trimi and Sheng, 2008). For example, it can augment the delivery of government information and services, foster equality by addressing the digital divide inherent in e-government services and applications. Moreover, it facilitates governments in diversifying their service channels and bolstering transparency, thereby mitigating issues such as corruption and low productivity.

In addition to the advantages offered by m-government, the adoption of new technologies and methodologies for m-government development also presents a series of challenges. Given the complex nature of governmental organizations, m-development challenges extend beyond mere technological issues, encompassing aspects of socio-technical issues, ranging from management, organization, policy, to legal matters, among others. It is noteworthy that these challenges are not exclusive to m-government but are also prevalent in the realm of e-government. A comprehensive review of the relevant literature has enabled us to categorize these challenges based on their relevance to the developing countries context. These challenges are shown in Table 2. It should be noted that while we have categorized each challenge under one primary category, it is possible that a challenge might also fall under another category. In addition, although these challenges are discussed in the context of developing economies, they may also be relevant in developed contexts. We have also compared the list of challenges with literature specific to Vietnam (e.g., White Book, 2019; Government Decision, 2019). We identified that Vietnam is confronting challenges similar to those discussed in the literature. Moreover, several challenges appear to be unique to Vietnam, as denoted by an asterisk (*) in Table 2.

**Table 2.** M-government challenges.

| No | The list of challenge | Challenge Category | Selected References |
|---|---|---|---|
| 1 | Lack of m-government regulations, laws in m-government deployment | Policy challenge | (Council, 2012) |
| 2 | Lack of integration standard regulations | | (OECD, 2011). |
| 3 | Lack of regulations for using m-government information | | (Al Thunibat, 2011) |
| 4 | Lack of strategy in m-government implementation | | (Maumbe et al., 2006; Sareen et al., 2013) |
| 5 | Lack of data sharing, protection regulations | | (Mengistu et al., 2009; Nguyen et al., 2015) |
| 6 | Lack of design principles* | | |
| 7 | Lack of security mechanisms for system security, Secure authentication, and access control | Security and privacy challenges | (Alkaabi, 2016; Goldstein , 2012; Marin, 2017) |
| 8 | Privacy violation | | (Sheng, 2008; Olanrewaju, 2013; Goldstein , 2012) |
| 9 | Lack of Security and privacy on mobile devices | | (Goldstein , 2012; Rannu, 2010) |
| 10 | Poor security on wireless infrastructure | | (Mengistu, 2009; Trimi, 2008) |
| 11 | Poor m-government infrastructure | Technology challenge | (Lee, 2006; Germanakos, 2005) |
| 12 | Limit of accessibility: Low bandwidth, low speed, low battery, and small screen | | (Kyem, 2016; Maumbe, 2006; Emmanouilidou, 2010) |
| 13 | Lack of interoperability between mobile applications | | (Lee, 2006; Hellström, 2008) |
| 14 | Lack of participation from end users and stakeholders | Organization and management challenge | (Mengistu, 2009) |
| 15 | Lack of cooperation mechanisms in m-government implementation | | (Goldstein , 2012; El-Kiki, 2007) |
| 16 | Lack of alignment between technological growth and organizational capacity | | (Goldstein,2012) |
| 17 | Lack of alignment of financial resources with investment in technology | | (Malik et al., 2013) |
| 18 | Poor in monitoring and evaluating the m-government implementation* | | |

### *2.3 Current status of e-government and m-government in Vietnam*

This research is aimed to the development of an EA framework for m-government in the context of a developing country, specifically Vietnam. This section thus reviews the current status of e-government and m-government in Vietnam.

The Vietnamese e-government has been developing for the past 20 years and is widely perceived as successful from the government's perspective, as reported in 2019 (White Book, 2019). The Vietnamese government has established a consolidated national public service portal (dichvucong.gov.vn) to deliver online public services to its customers. This portal is considered the primary channel for providing hundreds of online public services at the central government level. It is interconnected with other state agency portals, such as ministry-level online public service portals and provincial-level online service portals. However, the services offered on this national portal can currently only be accessed via personal computers and laptops. While information can be displayed on handheld mobile devices, the processing of services on these devices is not yet supported.

The government has enacted several policies to foster the development of e-government within the country. These include a Decree on data management, connection, and sharing, along with its guiding documents (Government Decree, 2020); a Decree on electronic identification and authentication for both individuals and organizations, accompanied by its guiding documents (Government Decree, 2022a); and a Decree on the protection of personal and organizational data, with its guiding documents (Government Decree, 2022b). In terms of the EA framework for e-government, Vietnam has promulgated a framework for e-government development (Government Decision, 2019). This framework serves as a crucial set of documents that guide government agencies in constructing information systems in line with EA principles. This approach helps to prevent redundant investments and reduce overall investment costs. In addition to the EA framework, the Vietnamese government has issued other policies related to the use of mobile applications in e-government. These include the management of e-commerce activities via mobile device applications (Government Circular, 2022 ); a pilot project implementing the use of telecommunications accounts (mobile money) for the payment of low-value goods and services (Government Decision, 2022); and regulations on the list of mandatory standards for digital signatures and digital signature authentication services, in accordance with the digital signature model on mobile devices and remote digital signing (Government Circular, 2019).

Vietnam has been also advancing its e-government towards m-government through the utilization of mobile applications (apps). The government has developed apps for various agencies to provide services to citizens and businesses. For example, the VSSID app, developed by the national insurance agency, offers social insurance services to residents and is integrated into the Vietnam national population database (Government VSSID, 2020). Another example is the Epoint EVN app, a mobile application from Vietnam’s electricity agency, which enables citizens to monitor their monthly electricity consumption and make payments (Government EVN, 2020). Moreover, numerous mobile applications for banking, e-wallets, e-payment, and mobile payment, such as PC-COVID (Government PC, 2021) and VNeID (Government VNeID, 2021), have been extensively adopted by government to manage and track movements during the COVID-19 pandemic.

## 3. Research Methods

### *3.1 Case context*

To provide the context of the study, we present crucial data pertaining to mobile phone usage in Vietnam. As depicted in Table 1 (White Book, 2019), there has been a significant increase in the number of mobile subscribers in recent years. The number of mobile phone subscribers surpasses the total population of Vietnam. A substantial majority of the population, specifically those over 18 years of age, possess a smartphone or similar device. This widespread adoption of mobile technology presents an advantageous opportunity for the Vietnamese government to advance its m-government initiatives.

**Table 3.** The number of mobile subscribers in Vietnam from 2016 to 2020.

| No | Items | Unit | 2016 | 2017 | 2018 | 2019 | 2020 |
|---|---|---|---|---|---|---|---|
| 1 | Number of mobile subscribers generates traffic (mobile phones and data cards) | subscriber | 128.996.179 | 120.016.181 | 136.088.885 | 132.429.054 | 129.454.026 |
| 2 | Number of mobile subscribers generating traffic per 100 people | % | 139,2 | 128,08 | 142,73 | 137.25 | 132.66 |
| 3 | Mobile subscriber number | subscriber | 125.454.516 | 115.014.658 | 130.385.371 | 126.150.541 | 123.626.427 |
| 3.1 | Active mobile phone subscribers using only voice and text messages | subscriber | 92.807.762 | 75.161.627 | 75.369.742 | 63.573.065 | 53.300.290 |
| 3.2 | Active mobile phone subscription with data usage | subscriber | 32.646.754 | 39.853.031 | 55.015.629 | 62.577.476 | 70.326.137 |
| 3.3 | Number of mobile phone subscribers per 100 people | % | 132,66 | 124,08 | 136,74 | 130.75 | 126.69 |

Moreover, 3G and 4G networks have been extensively deployed across the country (White Book, 2019). Specifically, 4G coverage extends to approximately 95% of the nation. In addition, 5G technology underwent trials and has been commercialized in 20 provinces and cities nationwide since 2020. These developments constitute favorable and essential prerequisites for the advancement of m-gov in Vietnam.

### *3.2 Research methods*

This research is designed in two phases. Phase 1 involves proposing a framework based on a literature review, the current EA frameworks, and the current state of e-government and m-government in Vietnam. Phase 2 evaluates the proposed framework. The two phases are described as follows:

In the first phase, we conducted a literature review on mobile government, considering the challenges of m-government. These challenges, which are likely to occur in Vietnam as well, include a lack of m-government policy for security, governance, and infrastructure development; a lack of mobile application infrastructure and integration; and the absence of an m-government implementation strategy. Each of these challenges is considered an input required for the m-government design. We then studied popular EA frameworks, including the Open Group Architecture Framework (TOGAF) (TOGAF, 2022), Pragmatic Enterprise Architecture Framework (PEAF) (PEAF, 2016), Federal

Enterprise Architecture Framework (FEAF) (FEAF, 2012), and Zachman Framework (Zachman, 2008), to determine the design approach for the EA framework for m-government. Next, we analyzed the context of e-government and m-government in Vietnam. The results of this phase are presented in Section 4.1, where we propose the EA framework for m-government by answering the following questions: What is the purpose of this framework? Who are the stakeholders in the framework? Why and when do they participate in framework design, and how do they interact? What are the framework components, and how can these components address m-government requirements?

In the second phase, we evaluated the proposed framework. The purpose of this phase is to predict user intentions regarding the proposed framework in practice (Mathieson, 1991). In particular, we chose the Ministry of Science and Technology (MoST) of Vietnam to conduct empirical data collection through a survey and poll of practitioners and e-government experts. The questionnaire was sent to the following informants: leaders and managers, data and applications specialists, IT infrastructure and security specialists, policy specialists, and finance specialists.

### *3.2.1 Theory of prediction user intentions*

Within the field of information systems (IS), the Technology Acceptance Model (TAM) (Davis, 1989) and the Theory of Planned Behavior (TPB) (Ajzen, 1991) are two widely used models for predicting a user’s intention to utilize IS artifacts (Mathieson, 1991). In this sense, the artifact in question is the proposed enterprise architecture. Unlike TAM, which provides general information about an individual’s opinion of an artifact, TPB offers more specific information that can better guide development (Mathieson, 1991). Also, the TPB model has been extensively utilized in numerous studies to understand human intentions across various contexts (Chen and Tung, 2014). Therefore, TPB was selected to evaluate and predict the likelihood of the proposed framework being utilized.

### *3.2.2 Survey design and implementation*

We employ a 5-point Likert scale to measure satisfaction for the the criteria that have been created. Interview and survey results are carefully reviewed, and non-responsive survey samples, such as those that do not answer all survey questions or respond illogically to the questionnaire, are excluded. Finally, the valid survey samples are analyzed and evaluated using the Likert scale to determine expert attitudes towards the applicability of the proposed framework.

In this study, we aim to understand the perceptions and attitudes of experts towards the proposed framework. The survey model is depicted in Figure 1. The criteria encompass addressing m-government’s challenges, the component architecture of the framework, and the quality of mobile apps. The experts’ opinions on these criteria will influence their attitudes towards the framework. If the experts disagree with the criteria, it is likely they will also disagree with the framework. After expressing their attitude towards the framework, experts will then indicate their intention to use the framework or not. The set of interview questions is provided in Appendix I

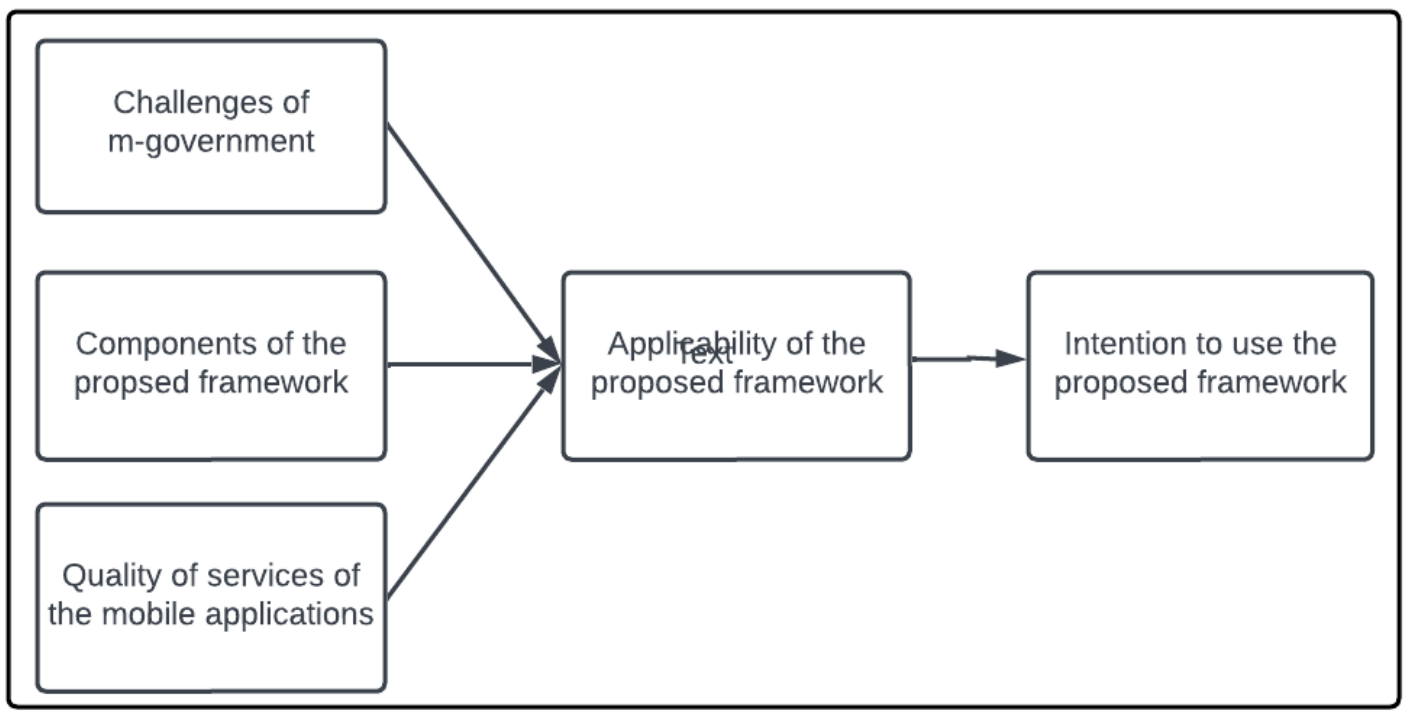

**Figure 1.** The survey model based on TPB

We employ the five-point Likert scale [55] to measure the consensus among experts. Specifically, we assigned a score of 1 to “strongly disagree” (SD), 2 to “disagree” (D), 3 to “neither” (N), 4 to “agree” (A), and 5 to “strongly agree” (SA). The collected data are analyzed as follows:

First, we calculate the total scores of the Likert scale by multiplying the frequency of each response option by its corresponding Likert scale score. Let’s denote $Total_s$ as the total score, $f_i$ as the frequency of each Likert scale score, and $i$ as the Likert scale score. We then have:

$Total_s = \sum(f_i \times i)$, where $i = \{SD, D, N, A, SA\}$ and $\{\mathrm{SD} = 1, \mathrm{D} = 2, \mathrm{N} = 3, \mathrm{A} = 4, \mathrm{SA} = 5\}$

Second, we calculate the mean scores of the Likert scale by dividing the total scores by the total number of respondents. Let’s denote $Mean_s$ as the mean score of Likert scale, $N_r$ is the number of respondents. We then have:

$$Mean_s = \frac{Total_s}{N_r}$$

Third, we interpret the range of the Likert scale mean score, which is assigned across three levels as follows: a score from 1.0 to 2.4 indicates a Negative attitude, a score from 2.5 to 3.4 indicates a Neutral attitude, and a score from 3.5 to 5.0 indicates a Positive attitude. The results of these calculations are illustrated in Appendix II.

In total, we conducted interviews with 26 experts and collected 26 completed questionnaires. However, one response was deemed invalid because the respondent selected “Strongly Disagree” (SD) for the criteria questions but chose “Agree” (A) for the attitude and behavior questions. Using an exclusion method, we removed this invalid response, resulting in 25 out of 26 valid responses for data analysis. Thus, the outcomes of the second phase are presented in Section 4.2

## 4. Findings

### *4.1 Proposed EA framework for M-government*

In this section, we introduce the proposed EA framework tailored for m-government. The framework, as depicted in Figure 2, is structured into three distinct phases, encompassing a total of eleven components. The detailed description of each phase and component is provided in the subsequent sections

#### *4.1.1 Initiation phase*

The initiation phase of an EA framework for m-government comprises three primary layers: the Viewpoint, Policy, and Design Principles. Each layer operates independently and delegates tasks to the subsequent layer. The layers are described as follows:

**Viewpoint Layer**: This layer presents various dimensions or perspectives that are relevant and beneficial to the designer, service provider, and users throughout the design process. It offers diverse perspectives on a to-be m-government, providing a comprehensive overview for the development of an m-government that satisfies requirements. The viewpoint layer addresses the following perspectives:

**- User Perspective**: Being integral to m-government, the user perspective is indicative of the success of the EA framework for m-government. It also takes into account factors such as mobile technology, technological literacy, the cost of utilizing mobile devices, and their service quality requirements. These elements form a set of requirements that EA designers need to thoroughly address.

**- Finance Perspective**: From the financiers' viewpoint, it should be able to demonstrate the effectiveness of adopting EA framework for m-government, aligning investment objectives with the efficiency of the administration. The financial perspective also aids in balancing financial resources to meet investment projects during the next phases.

**- Technical Perspective**: This perspective represents the views of professionals, technicians, and information technology experts. It will display system interoperability among ISs of state agencies.

**- Management Perspective**: This perspective represents the views of managers when providing public services to customers. It illustrates the benefits of implementing EA for m-government, from which they formulate policies to promote, manage, and operate m-government, and construct a roadmap to transition from e-government to m-government.

**- Vendor Perspective**: This perspective represents the views of network operators and value-added service providers such as electronic payment applications, express delivery, etc. Their involvement aids in building ICT infrastructure and logistics services for m-government.

**Policy Layer:** The policy serves as the foundation for the EA framework in the context of m-government. m-government is perceived as a subset of e-government. Thus, policies pertaining to m-government fall under the broader scope of e-government policies (Ntaliani et al.,2008). It is proposed that the EA framework is more effective when it possesses the capability to be implemented and enforced. This implies its utilization within an efficient and effective legal and regulatory context (McMillan, 2010), thereby facilitating service provision through m-government (Germanakos et al., 2005).

**Design Principles Layer:** The design principles are employed to guide stakeholders in addressing issues that organizations are likely to encounter (Bharosa et al., 2011). These principles are recognized as common rules and guidelines in the development of EA, aiding organizations in fulfilling their mission (TOGAF, 2022). Depending on the organization, these principles could be established across different dimensions and at various levels. The design principles in this framework aim to meet EA design perspectives by mandating EA designers to adhere to design principles for data, applications, and integration.

### *4.1.2 Design and implementation phase*

There is no global agreement on EA framework layers itself. For example, National Institute of Standards and Technology (NIST) EA model, there are five layers, namely business architecture, information architecture, information systems architecture, data architecture, and delivery systems architecture (NIST, 1990). In a similar vein, Federal Enterprise Architecture Framework (FEAF) has four layers, that is, business architecture, data architecture, application architecture, and technology architecture (FEAF, 2012). TOGAF ADM, on the other hand, presents three main architectures, including business architecture, information system architecture, and technology architecture (TOGAF, 2022). We thus propose EA framework by keeping most prominent layers that have appeared in popular framework, such as business architecture, data architecture, application architecture. Taking into consideration of m-government is designed to integrate into e-government, we stress the importance of integration by dedicating a layer for that i.e. integration architecture. The integration is often a part of technology architecture in popular EA frameworks. Similarly, security is considered as a part of technology architecture as in FEAF and TOGAF ADM, or in delivery systems architecture in NIST EA model. Here due to the importance of digital technologies in EA for m-government, a security architecture layer is proposed to stress the importance of this perspective in m-government. As a result, there are six layers and their details are presented as follows:

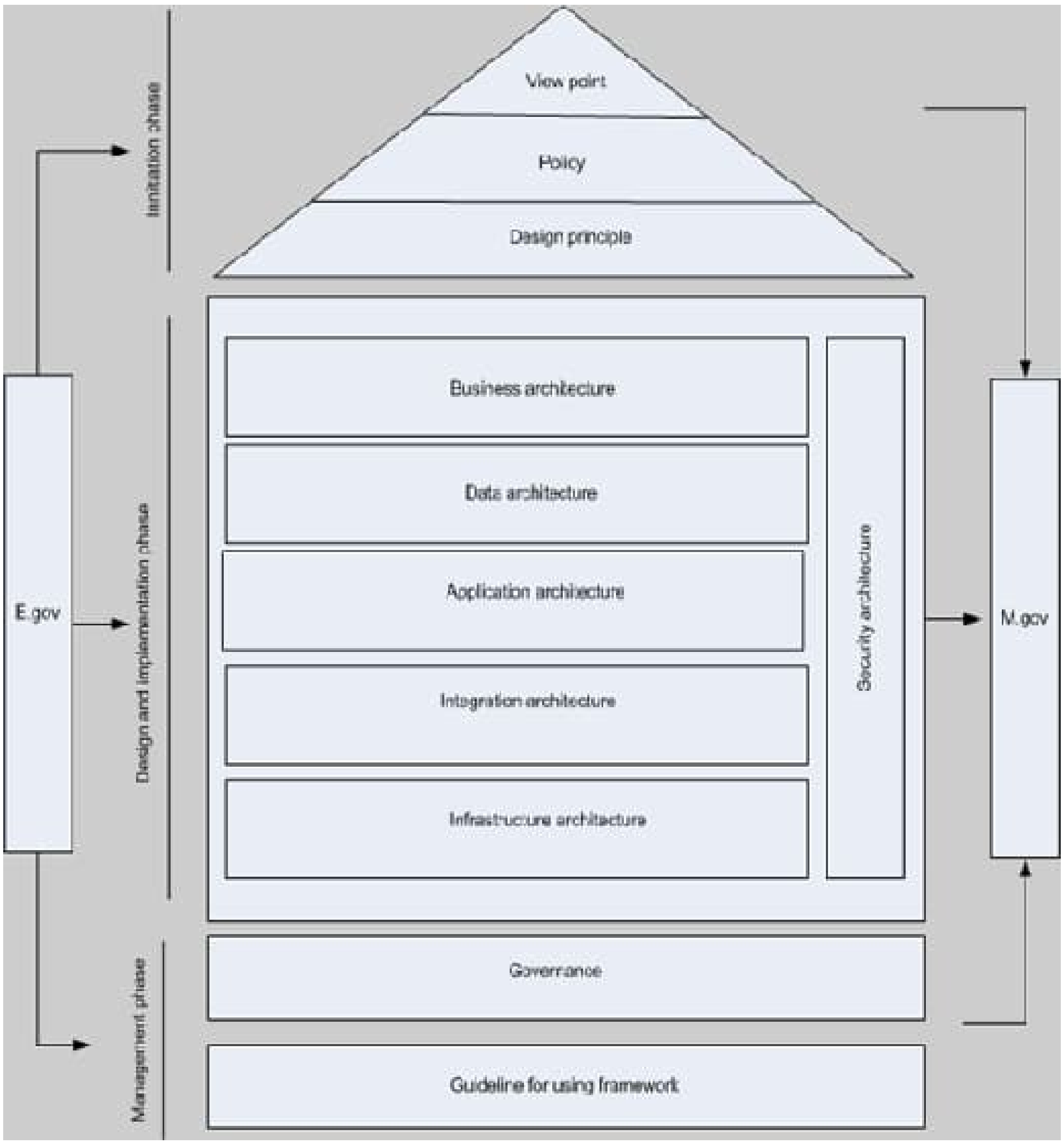

**Figure 2.** Enterprise architecture framework for m-government

**Business architecture.** This model describes the business processes that take place within government agencies and the business processes of public services provided to citizens and businesses. This architecture guides other architectures, such as data architecture, application architecture and security architecture.

**Data architecture.** The data architecture describes the data needed for business processes. Data for m-government inherits from e-government. However, there will be other data sources generated from devices such as mobile phones, IoT, and GIS (Julsrud and Krogstad, 2020). Data from m-government can come from several sources, such as GPS, Bluetooth, Wi-Fi positioning, and motion systems (Wang et al., 2018; Panduranga et al., 2020), social media, sensor networks, and volunteered geographic information (Kafi et al., 2012; Brynielsson et al., 2018).

**Application Architecture.** Similar to e-government, m-government provides services such as government to citizen (G2C), government to business (G2B), government to government (G2G), and government to employee (G2E) (Ntaliani et al., 2008; Mustafa and Shabani, 2018). M-government can offer multiple channels, such as SMS, unstructured supplementary service data (USSD), mobile web, mobile application, and voice channels (Isagah & Wimmer, 2019). These channels used to aid other businesses, such as monitoring car-GPS tracking by insurance services (Derikx et al., 2019), ride-sharing uses GPS location data (Aïvodji et al., 2016), and privacy data for energy services (Grünewald and Reisch, 2020).

**Integration Architecture**. M-government is built on e-government (Rahmadany and Ahmad, 2021) by adopting mobile technologies (Germanakos et al., 2005). Therefore, it is important to integrate m-government into existing e-government. Integration of m-government to e-government includes application, security, standard of the data, application program interface (API) (Isagah and Wimmer, 2018).

**Security Architecture**. M-government has faced a high risk of cybersecurity due to ubiquitous connectivity of technologies such as 3G, 4G, 5G, Wi-Fi, Bluetooth, and wired connections (Harvey et al., 2014). This creates concerns for citizens and businesses when participating in government public services through their mobile devices. Therefore, this layer ensures a holistic approach for securing government services provided through m-government (Harvey et al., 2014; Bahar et al., 2013).

**Infrastructure Architecture**. M-government supplements e-government (Kumar and Sinha, 2007) and m-government must be based on e-government infrastructure (Marin et al., 2017). Example of m-government infrastructure include 3G, 4G, 5G networks, and mobile IPv6. These infrastructures reinforce smooth streaming of video, and highly secured data transmission (Kumar et al., 2016; Foghlú, 2005).

### *4.1.3 Management phase*

This phase performs the tasks of managing and monitoring stakeholders when participating in the implementation and operation of the EA framework, monitoring the implementation road map of the components in the framework, and giving guidelines for using the EA framework to help stakeholders to coordinate smoothly when implementing the framework. There are two layers of this phase as follows.

**Governance Layer**. To effectively manage the EA framework, the government needs to develop a road map for implementation and clearly define the tasks for the government agencies involved in developing the components of the framework, such as the duties of network operators, public service providers, or financial providers. The government also needs to manage and operate the implementation of that roadmap. Require stakeholders to commit to performing and completing tasks on schedule. At the same time, conduct an assessment of the effectiveness of the EA framework, and review the shortcomings and challenges that need to be addressed. In sum, the roadmap help us to answer the question who is doing what, two broad communities use the EA: architects and stakeholders. Stakeholders include decision-makers, and implementers. Each of these communities uses the architecture differently (Lankhorst, 2009).

**Guideline Layer**: The Guideline is not mandatory but it is essential in guiding stakeholders as they participate in the EA design process. It is up to the structure of the organization and the applicable environment to provide an appropriate guideline. Each stakeholder has a different role in EA design. Therefore, the guideline helps stakeholders define their roles and tasks, clearly seeing cooperation with other stakeholders. Since there are many stakeholders involved in the design process, and from the complexity of the relationships between stakeholders and the need for handling that arise from the users, in this framework we recommend that the guide be used in government agencies to support the design process.

### *4.2 Predicting user intentions for the proposed EA framework in m-government*

In this section, we present the results of the data analysis conducted to predict user intentions for the proposed framework, based on the methodology described in Section 3.2.

The first group of criteria, consisting of five questions addressing m-government's challenges, is numbered from I.1 to I.5. The frequency of the rating levels for each question is summarized in Table 4. Subsequently, an evaluation chart is created as shown in Figure 3. The chart reveals that no expert chose the "Strongly Disagree" (SD) level, about 8% of respondents chose "Disagree" (D), nearly 15% of respondents expressed a neutral attitude, about 50% of respondents chose "Agree" (A), and the remaining 45% of respondents chose "Strongly Agree" (SA). Referring to Table AII.1 of Appendix II, the results of calculating the mean score of each criterion and the overall mean score are also positive. This suggests that the first criterion has attracted a positive attitude from experts.

**Table 4.** The frequency of scale in the first criteria

| Scale \ Criteria | I.1 | I.2 | I.3 | I.4 | I.5 |
|---|---|---|---|---|---|
| Strongly disagree | 0% | 0% | 0% | 0% | 0% |
| Disagree | 8% | 8% | 8% | 8% | 8% |
| Neither | 8% | 8% | 24% | 4% | 4% |
| Agree | 52% | 56% | 40% | 56% | 28% |
| Strongly agree | 32% | 28% | 28% | 32% | 60% |

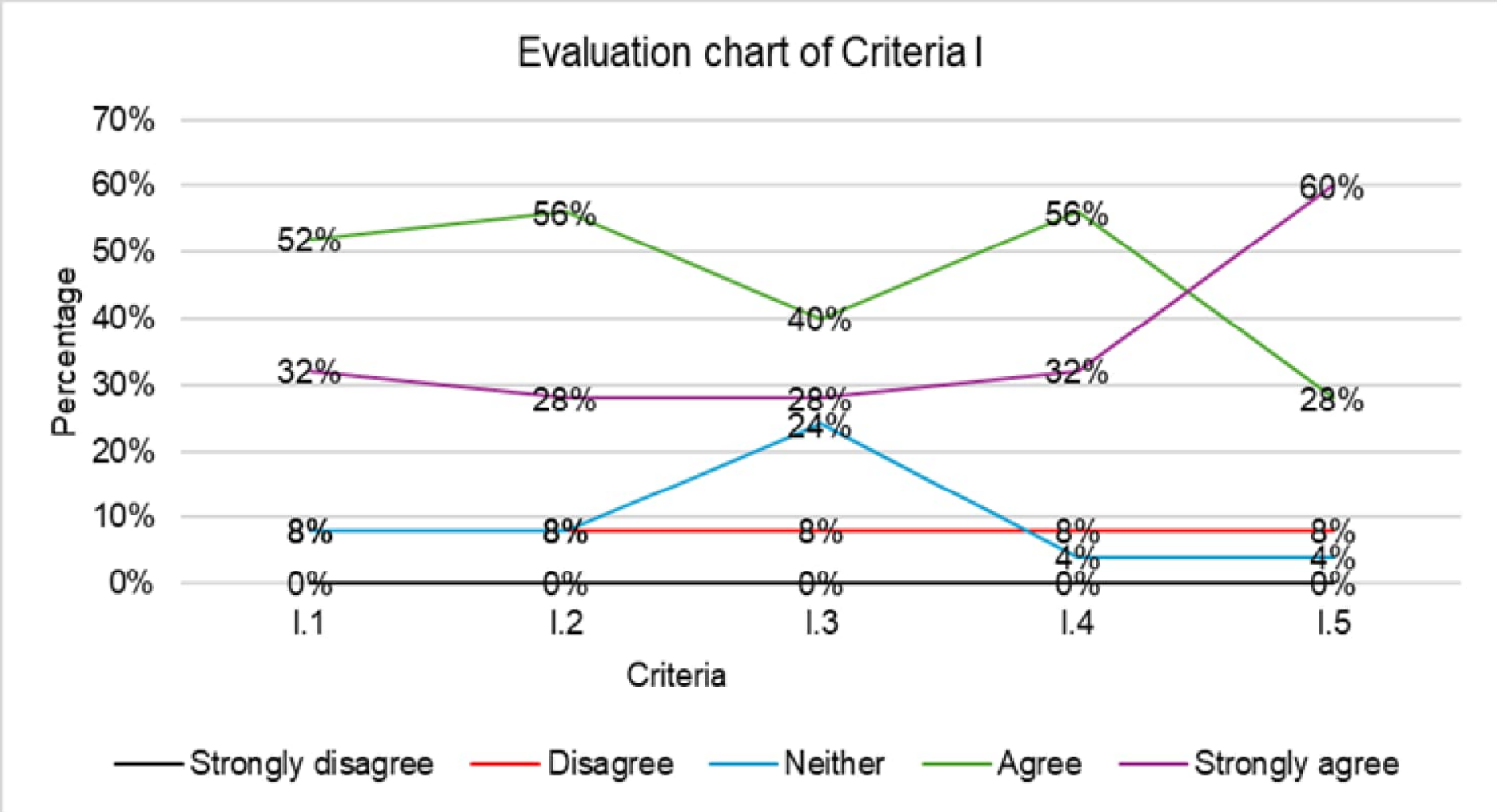

**Figure 3.** The evaluation chart of addressing M-government's challenges

In the second group of criteria, there are six questions related to the component architecture of the framework. The frequency of their scale is summarized in Table 5. The evaluation chart in Figure 4 reveals that less than 10% of respondents expressed attitudes at the "SD", "D", and "N" levels. The remaining 40% of respondents chose "A", and 45% of respondents selected "SA". This indicates that the majority of experts concur with the choice of Business Architecture, Application Architecture, Data Architecture, Integration Architecture, Infrastructure Architecture, and Security Architecture as the component architectures of the framework. This result aligns with the positive attitude calculated in Table AII.2, Appendix II.

**Table 5.** The frequency of scale in the second criteria

| Scale \ Criteria | II.1 | II.2 | II.3 | II.4 | II.5 | II.6 |
|---|---|---|---|---|---|---|
| Strongly disagree | 0% | 0% | 0% | 0% | 0% | 0% |
| Disagree | 8% | 8% | 8% | 8% | 8% | 8% |
| Neither | 4% | 4% | 4% | 4% | 4% | 8% |
| Agree | 44% | 48% | 44% | 36% | 36% | 44% |
| Strongly agree | 44% | 40% | 44% | 52% | 52% | 40% |

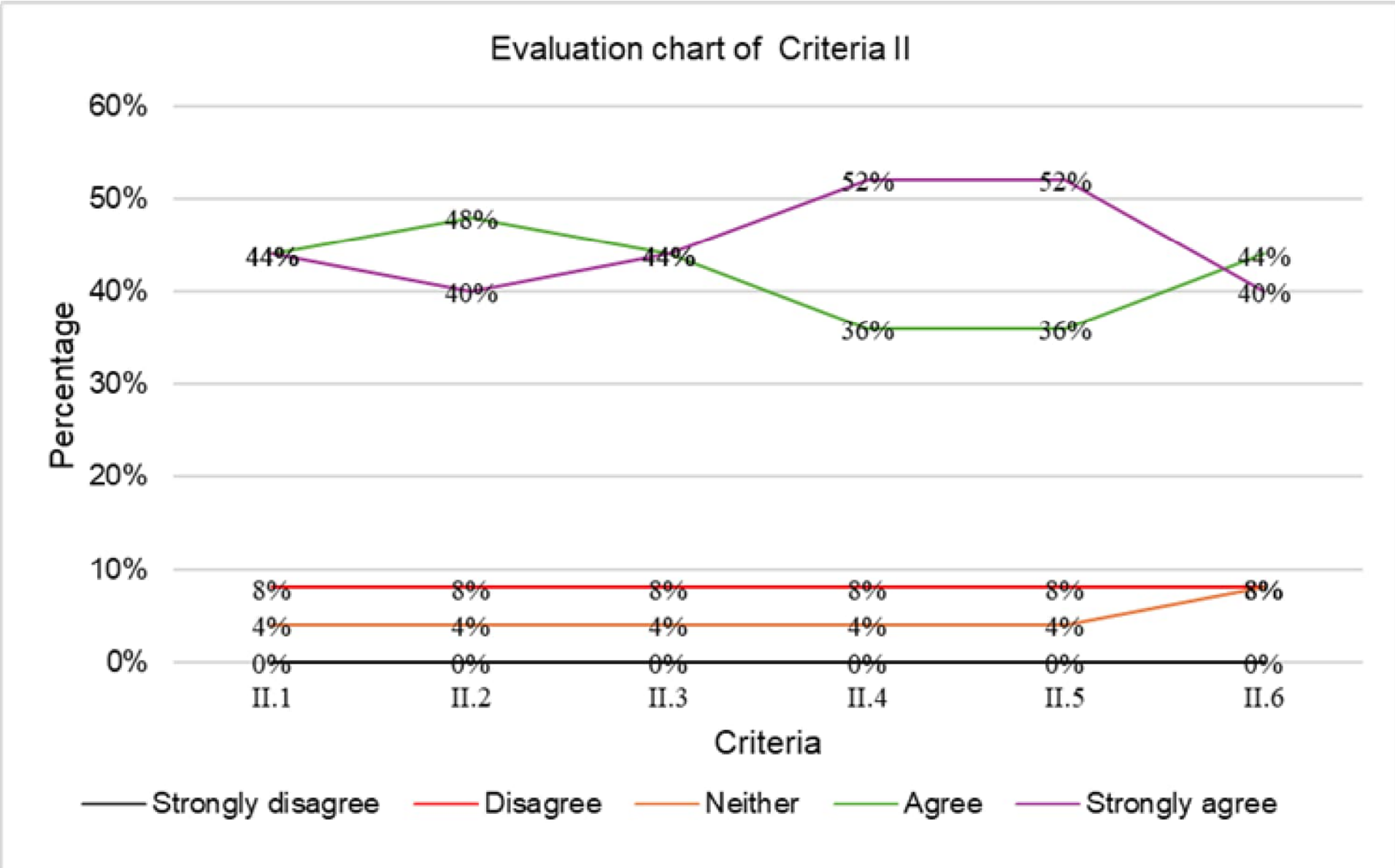

**Figure 4.** The evaluation chart of framework component architecture

The third group of criteria pertains to the quality of service of the mobile application, the frequency of which is summarized in Table 6. The evaluation chart depicted in Figure 5 reveals that a mere 10% of experts provided comments at the N level, less than 10% opted for D, and none selected SD. Approximately 45% of experts chose either A or SA. Notably, nearly 60% of experts concurred with the proposition that there should be coordination among stakeholders in the application of mobile services. A similar percentage expressed strong agreement with the necessity of integrating mobile payment applications into mobile applications. Reference to the results presented in Table AII.3 of Appendix II also yields a positive attitude. Consequently, the majority of the criteria for constructing the m-government framework have received substantial agreement from experts.

**Table 6.** The frequency of scale in the third criteria

| Criteria / Scale | III.1 | III.2 | III.3 | III.4 |
|---|---|---|---|---|
| Strongly disagree | 0% | 0% | 0% | 0% |
| Disagree | 4% | 8% | 4% | 4% |
| Neither | 8% | 12% | 12% | 8% |
| Agree | 40% | 48% | 56% | 32% |
| Strongly agree | 48% | 32% | 28% | 56% |

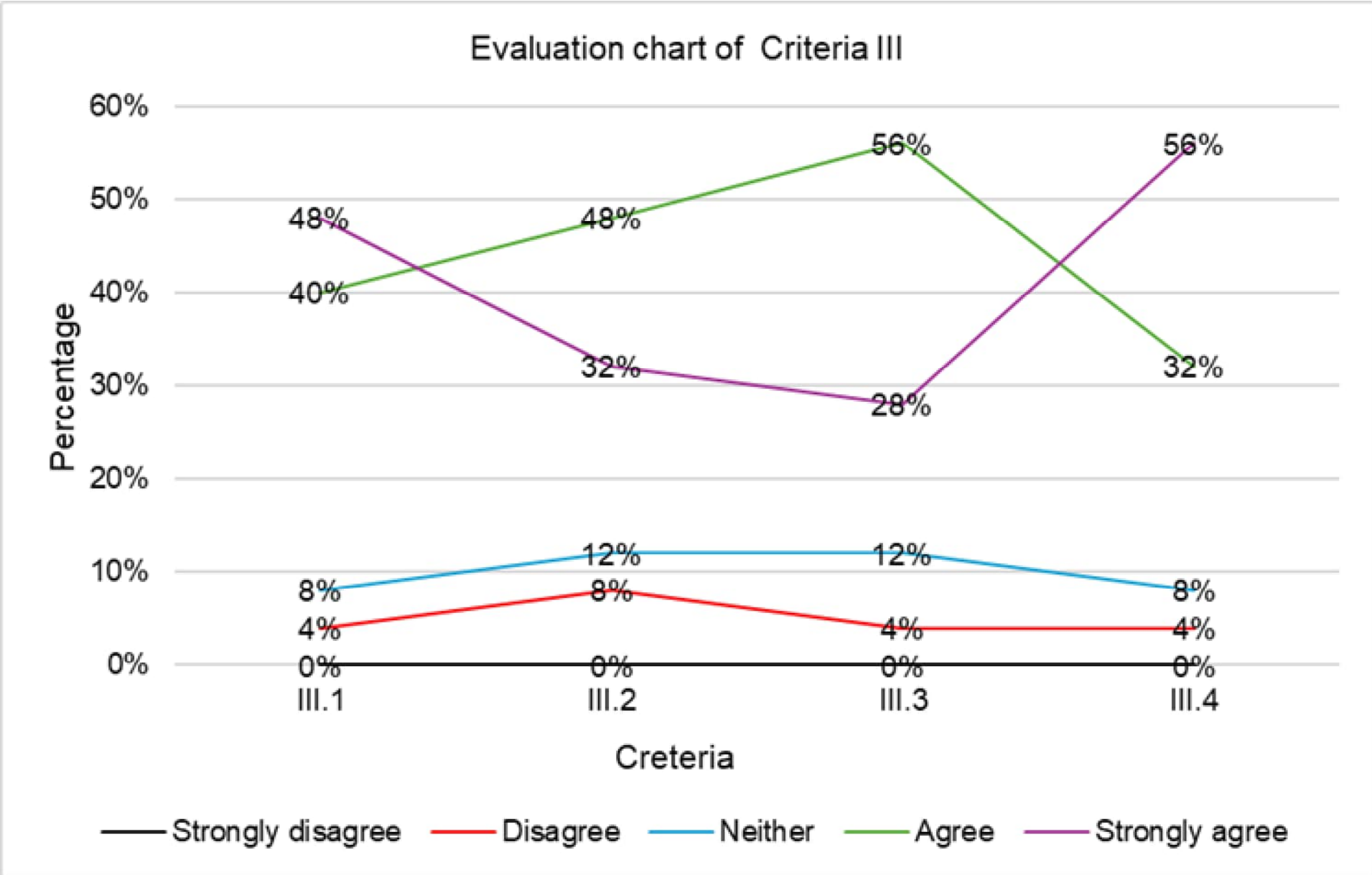

**Figure 5.** To ensure of the quality of service of the mobile application

Expert opinions on the criteria for developing the framework have been gathered to assess its applicability at the MoST, as summarized in Table 7. Figure 6 illustrates that 58% of experts concur and 12% strongly concur with the applicability of the framework. Approximately 25% of experts maintain a neutral stance, with a small minority of around 7% expressing disagreement and none strongly disagreeing. The calculations presented in Table AII.4 of Appendix II indicate that the mean scores of attitudes range from 3.7 to 3.8, suggesting a positive attitude towards the applicability of this framework among the experts.

**Table 7**. Frequency of scale in expert's applicability assessment

| Scale \ Criteria | IV.1 | IV.2 | IV.3 | IV.4 |
|---|---|---|---|---|
| Strongly disagree | 0% | 0% | 0% | 0% |
| Disagree | 4% | 4% | 8% | 12% |
| Neither | 28% | 28% | 24% | 20% |
| Agree | 56% | 56% | 56% | 56% |
| Strongly agree | 12% | 12% | 12% | 12% |

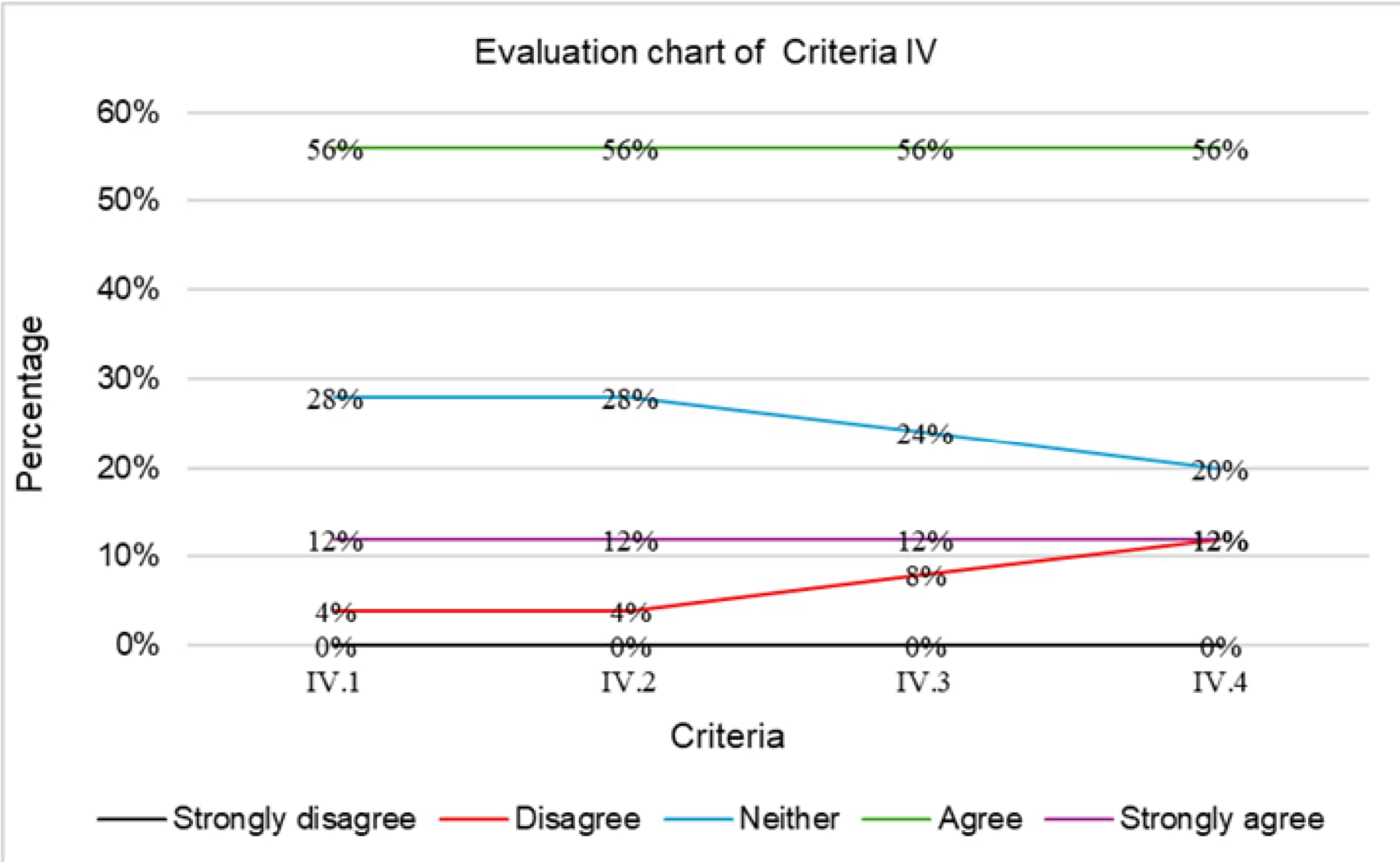

**Figure 6.** The evaluation chart of framework applicability assessment

We propose three potential scenarios for the application of this framework at the MoST. These scenarios include immediate application, application following the preparation of investment funds, and application upon readiness to provide public services via the mobile application. The results of these proposals are summarized in Table 8, and an evaluation chart is presented in Figure 7. The data reveals that approximately 60% of experts agree, and 15% strongly agree, with the proposition for the MoST to utilize this framework. Conversely, about 22% of experts remain neutral, 15% disagree, and none strongly disagree. These findings align with the positive attitude calculated in Table AII.5 of Appendix II.

**Table 8.** Frequency of scale in framework selection intention

| Criteria / Scale | V.1 | V.2 | V.3 |
|---|---|---|---|
| Strongly disagree | 0% | 0% | 0% |
| Disagree | 8% | 8% | 8% |
| Neither | 32% | 24% | 16% |
| Agree | 48% | 56% | 72% |
| Strongly agree | 12% | 12% | 4% |

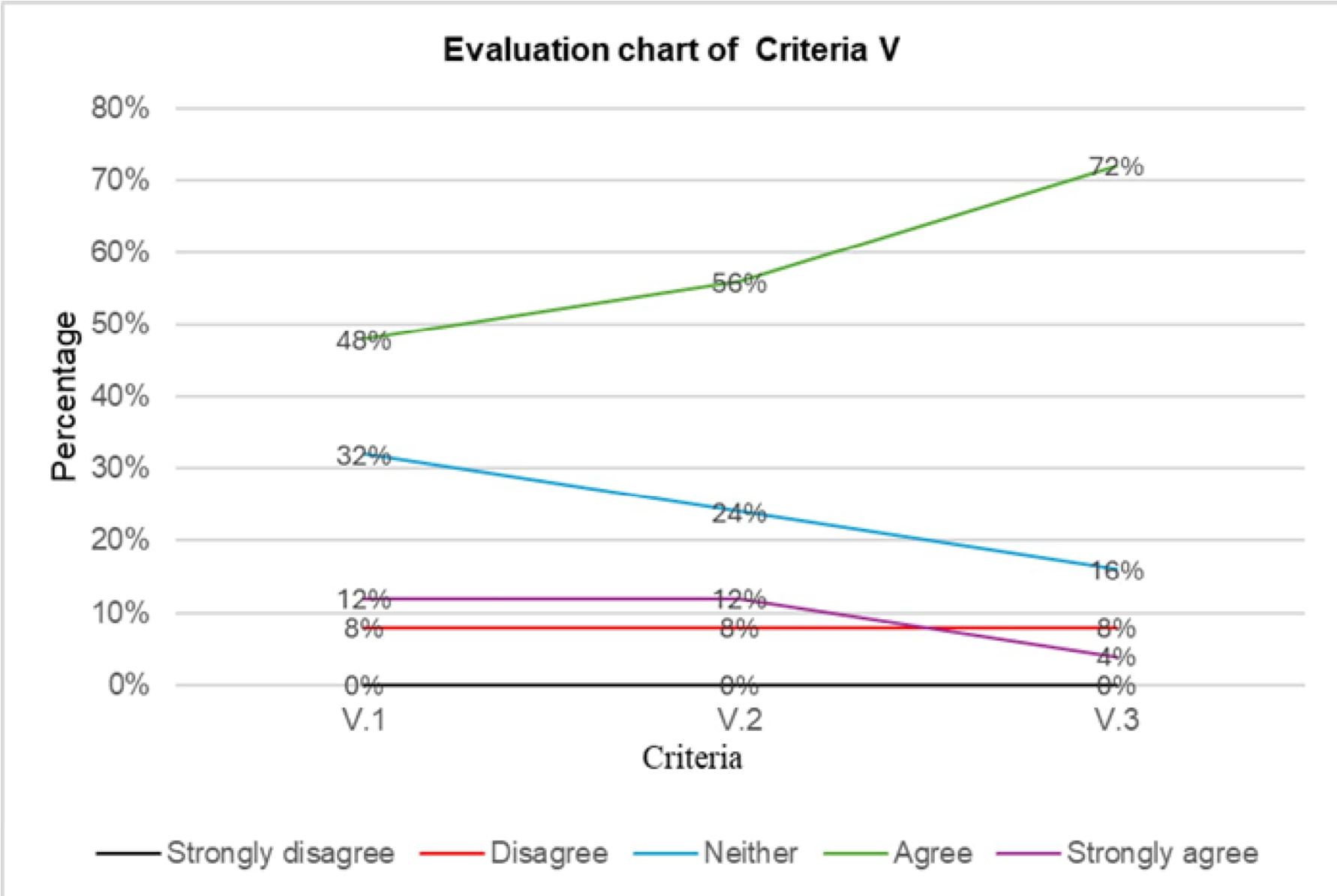

**Figure 7.** The evaluation chart of framework selection intention

## 5. Discussion

### *5.1 Compatibility of the proposed framework*

Several countries, including Bangladesh, Oman, Jordan, South Africa, and Malaysia, have developed and implemented m-government frameworks (Joshi et al., 2015). To assess the compatibility of our proposed framework with others, we selected 12 related m-government frameworks from countries with a context similar to Vietnam, i.e., developing countries.

As depicted in Table 9, the majority of these frameworks address technological challenges, encompassing data, applications, and network infrastructure. Significant consideration is also given to stakeholder collaboration and business process issues, with 10 out of 12 frameworks addressing these two issues. In addition, these frameworks are applicable for addressing policy and management issues, with seven out of 12 frameworks considering these two issues as primary concerns. Only a quarter of the frameworks pay attention to security, a common concern in developed countries. Other issues such as finance and integration receive minimal attention, with no framework addressing integration.

In conclusion, a comparative analysis of the proposed framework with existing ones reveals that our framework encompasses all the criteria outlined by the others. Additionally, it introduces a novel criterion, namely, integration, thereby enhancing the comprehensiveness of the existing frameworks. As a result, it can be inferred that our framework, while being particularly suitable for Vietnam, also exhibits a high degree of compatibility with other countries. By addressing all the criteria, our proposed framework provides a distinct advantage and stands as a leading example in mobile government EA, thereby effectively addressing the challenges of mobile government.

**Table 9.** The comparison of m-government frameworks in developing countries

| Framework / Criteria | Proposed framework | Spoke and hub modle (Narayan, 2007) | MERS framework (Amailef, 2008) | Iran framework (Fasanghari, 2009). | Management framework (El-Kiki et al, 2005) | Jordan framework (Al-Masaeed, 2013) | Oman framework (Al-Hadidi, 2010) | Livelihood framework (Duncombe, 2012) | Agent Based Framework (Al-Sakran et al., 2013) | Africa framework (Henning et al., 2019) | Bangladesh framework (Hossain et al., 2015) | South Korea framewor (Kim, et al 2004) | Mongolia framework (Erdenebold, 2014). |
|---|---|---|---|---|---|---|---|---|---|---|---|---|---|
| Developing country context | x | x | x | x | x | x | x | x | x | x | x | x | x |
| Policy issues | x | | | x | | x | x | | x | x | x | | |
| Stakeholder issues | x | | x | | | x | x | x | | x | x | | x |
| Business progress issues | x | x | | x | x | x | | | x | x | x | x | x |
| Technology issues (data, apps and infrastructure) | x | x | x | x | | x | x | x | x | x | x | x | x |
| Security issues | x | | | | | | x | x | x | x | x | | x |
| Integration issues | x | | | | | | | | | | | | |
| Financial issues | x | | | | | | x | | | | x | | |
| Governance issues | x | x | | x | x | x | x | | x | x | x | | |
| Strategy issues | x | | | x | | | x | | | x | x | | |

## *5.2 Applicability of the proposed framework in the context of Vietnam*

In addition to the positive attitudes from experts towards the proposed framework, as indicated by the results of the prediction of user intentions presented in the aforementioned section, we also conducted a comprehensive analysis of the framework's alignment with the current development of e-government, its infrastructure, and policy. This suggests a high likelihood of user intent to utilize the framework for several reasons.

First, Vietnam is currently implementing a digital transformation strategy aimed at transitioning towards a digital government (Government Decision, 2020). One of the key pillars of Vietnam's digital transformation initiatives is the provision of services based on smart devices. Consequently, it is likely that the existing e-government will transition to m-government. During this transformation process, our proposed framework can serve as a valuable reference for the government and state agencies. It addresses the challenges and provides recommendations for establishing a robust m-government.

Second, according to a survey conducted by the Ministry of Information and Communications of Vietnam, which oversees IT applications in state agencies and digital transformation initiatives in the country (Report, 2023), state agencies are encountering difficulties in consolidating common views from various stakeholders in the implementation of EA. While existing frameworks or methods such as TOGAF or FEA are available (TOGAF, 2022; FEAF, 2012), the proposed framework offers an alternative approach that could assist state agencies in reconciling conflicting viewpoints. For instance, the framework advocates for the consideration of perspectives from diverse stakeholders, including managers, finance, engineering, service providers, and users. By harmonizing these perspectives, it facilitates state agencies in achieving greater success in establishing their mobile government framework (Dang and Pekkola, 2020; Do et al. 2023).

Third, given the existence of an EA for e-government in Vietnam (Government Decision, 2024), the proposed framework presents an opportunity for transitioning from the current EA for e-government to an EA for m-government. This transition is feasible due to our framework's ability to inherit components from the e-government EA. For instance, the application architecture of the proposed framework incorporates components from the existing EA for e-government applications, as well as new features specific to mobile devices, such as location-based services, cameras, GPS, and big data warehouses. In addition, it includes analysis tools for handling unstructured and semi-structured data generated by mobile devices.

Fourth, based on the national digital transformation strategy for the period of 2021-2025, with a vision extending to 2030 (Government Decision, 2020), the development of frameworks within state agencies is identified as a key task to achieve the government's goal of digital governance. However, the current administrative structure in the country is organized into two distinct levels: the central government and the local government. This dichotomy often results in miscommunications when developing a comprehensive framework that requires the involvement of various stakeholders across different agencies (Report, 2023). To address this, we propose a roadmap, as depicted in Figure 8, alongside our framework. This roadmap aims to operationalize the national digital transformation strategy for the period of 2021-2025 and facilitate the implementation of our framework.

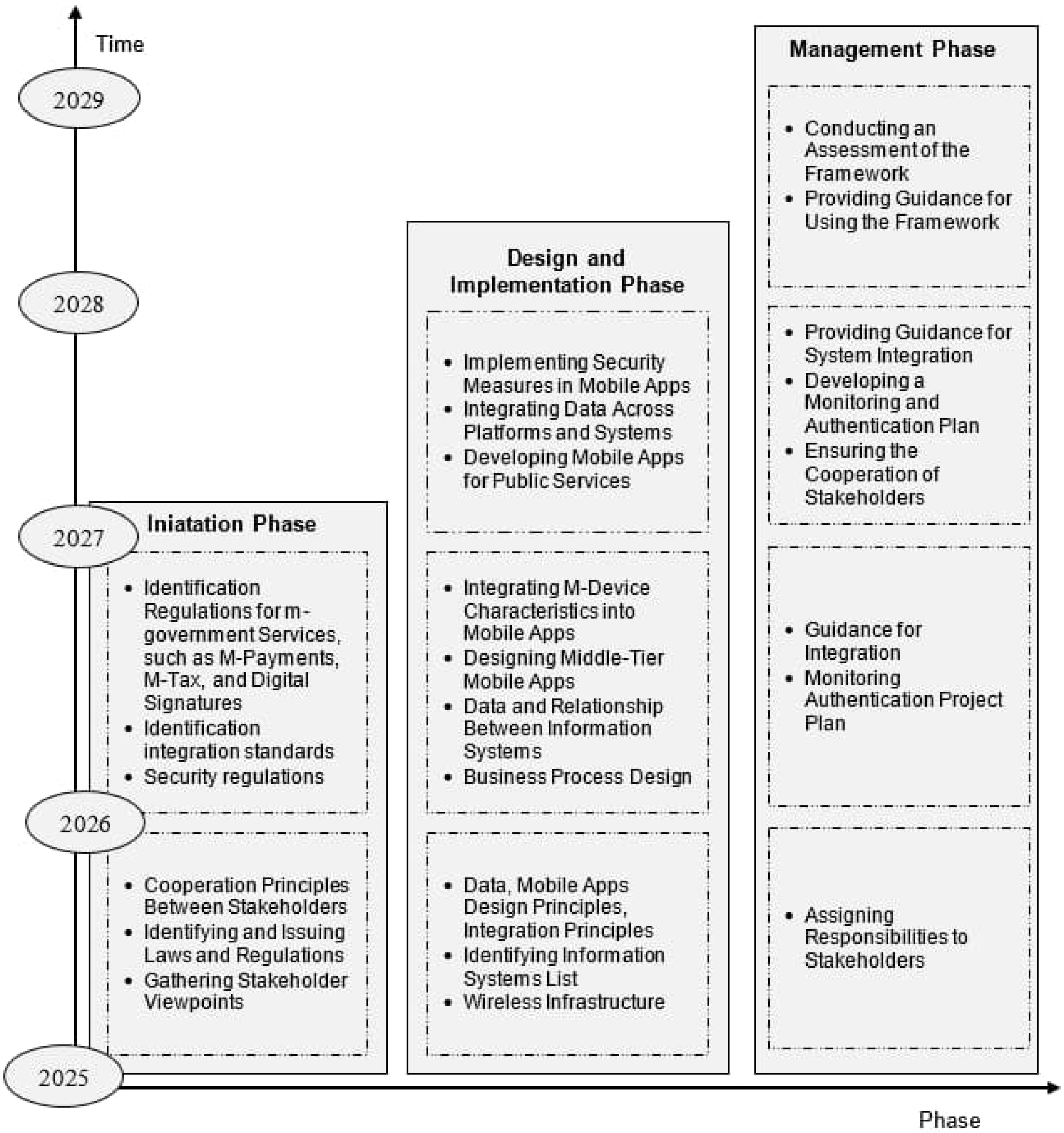

**Figure 8.** The m-government management roadmap.

Finally, there are fundamental causes for the adoption of EA in the public sector (Dang and Pekkola, 2016). One such cause is the lack of implementation policies. In this context, Vietnam's m-government policies, which are currently based on e-government policy, are still in their infancy. The proposed framework is inline with that by suggest that the Vietnamese government should establish legal regulations at the initial stage of the m-government EA framework development. This would create a legal foundation before mobile applications are introduced in the m-government during the subsequent implementation phase. Such policies could include, for instance, the use of mobile phone numbers as identifier codes for m-government applications (such as m-vote, m-ticket, etc.), and regulations to safeguard privacy when using mobile devices.

## 6. Conclusion

This study presents a framework for EA in mobile government, encompassing three primary phases: the initiation phase, the design and implementation phase, and the management phase. Each phase is further delineated into its constituent layers or components. The proposed framework has received positive feedback regarding user intentions. Moreover, it demonstrates compatibility with other m-government frameworks and applicability within the context of Vietnam.

The study also offers practical implications. First, for practitioners, the framework provides a methodology to establish an m-government framework for state agencies by adhering to the outlined phases, their components, and layers. Second, practitioners can use this framework as a guide to transition from the existing EA for e-government to EA for m-government. It also assists in addressing common challenges in establishing EA, thereby potentially reducing costs by minimizing duplicate investments.

**Limitation.**

There are several limitations. First, the framework was established based on literature and tested in terms of predicting user intentions. In that sense, we used an approach similar to Akhlaghpour and Lapointe (2018), utilizing existing academic and practitioners' literature to develop a framework based on reasoning (Mantere & Ketokivi, 2013). Although it received a positive response towards intentions to use, it remains a theoretical model. Therefore, it necessitates pilot testing within an organization for evaluation. We are currently in the process of testing our framework and adjusting it using design science (Hevner et al., 2004) to further develop and evaluate the framework. In addition, we are employing citizen science (Mäkipää et al. 2020; Dang et al., 2022) as a mechanism to gather public feedback, which is crucial in the development of the framework.

Second, the adoption of TPB often implies the usage of Partial Least Squares (PLS) methods. However, due to the insufficient number of experts conducting EA within the case study (e.g., MoST), we were unable to perform PLS (Goodhue et al., 2012). Consequently, we had to analyze the sample by means of calculation. In the future, we plan to gather more respondents so that we can apply PLS to our data.

Third, we were aware that the respondents may not be familiar with crucial TPB concepts, such as attitude and intention. The lack of clear differentiation between these significant concepts could have caused difficulties for respondents in providing accurate responses. To address this, we tailored the TPB for the study context and provided carefully worded explanations in the respondents' preferred languages. We hope this approach may have helped to minimize inaccuracies in responses and thus improve the reliability of the results.

Fourth, the TPB was used to predict user intentions. We acknowledge that this theory may not be suitable for evaluating the applicability of the proposed framework. To address this, we conducted a comprehensive analysis of the framework's alignment with the current development of e-government, its infrastructure, and policy in the study context. In the future, this framework will be evaluated for its applicability in conjunction with design science, citizen science, and qualitative case study.

## APPENDIX I: THE INTERVIEW QUESTIONNAIRE

**Table AI.1.** Expert site

| **Expert site** | **Percentage** |
|---|---|
| Ministry of Science and Technology | 78.6% |
| Ministry of Information and Communication | 21.4% |

**Table AI.2.** The Expert's field

| **Expert's field** | **Percentage** |
|---|---|
| Leader and manager | 21.4% |
| application and data specialist | 28.6% |
| IT infrastructure and security specialist | 14.3% |
| Policy specialist | 25% |
| Finance and budgeting Specialist | 10.7% |

**Table AI.3.** Assess the importance of public service delivery on mobile applications

| **Likert scale** | **Percentage** |
|---|---|
| Very important | 60.7% |
| Important | 25% |
| Normal | 14.3% |
| Not important | 0% |
| Very unimportant | 0% |

**Table AI.4**. The interview questionnaire

| **Question number** | **Item** |
|---|---|
| **I** | **The criteria of addressing the challenges of M-government** |
| I.1 | Develop policies before operating mobile applications (Council, 2012). |
| I.2 | Provide guidelines and management road maps (Al Thunibat et al., 2011; Maumbe and Owei, 2006 ) |
| I.3 | Align technology investment results with the organization's goals and vision (Olanrewaju, 2013). |
| I.4 | Provide mechanisms to ensure information security (Goldstein et al., 2012) |
| I.5 | Provide design principles to avoid duplication of investment (Narayan, 2007). |
| **II** | **The criteria of the framework's component architectures** |
| II.1 | Business Architecture (TOGAF, 2022; FEAF, 2012) |
| II.2 | Application Architecture (TOGAF, 2022; FEAF, 2012) |
| II.3 | Data Architecture (TOGAF, 2022; FEAF, 2012) |
| II.4 | Integrated Architecture (Mengistu et al., 2009) |
| II.5 | Infrastructure architecture (FEAF, 2012) |
| II.6 | Security Architecture (FEAF, 2012) |
| **III.** | **The criteria for mobile apps service quality** |
| III.1 | Convenient, guaranteed service 24/7, anytime, anywhere (Mengistu et al., 2009) |
| III.2 | Gather opinions, including user opinions, to provide a comprehensive service (Mengistu et al., 2009) |
| III.3 | Ensure the cooperation of the stakeholders to provide service smoothly (Hellström, 2008). |
| III.4 | Integrate with mobile payment apps to make sure convenient service (Queensland government, 2018) |
| **IV** | **Assess the applicability of the framework** |
| IV.1 | This framework is consistent with the Ministry's public service delivery model |
| IV.2 | This framework has the capability to make sure the Ministry avoids duplicate |
| | investments |
| IV.3 | This framework has the capability to make sure the Ministry to deliver a mobile application ecosystem |
| IV.4 | This framework helps the Ministry to have a clear roadmap for building and managing mobile applications |
| **V** | **Framework selection intention** |
| V.1 | I would suggest adopting this framework right now |
| V.2 | I will consider the application when the Ministry ensures a sufficient investment budget |
| V.3 | I will propose to apply this framework when the Ministry is oriented to provide public services on mobile applications |

## APPENDIX II: DATA ANALYSIS RESULTS

**Table AII.1**. Data analysis results of criteria I

| Option | Scale | I.1 | I.2 | I.3 | I.4 | I.5 |
|---|---|---|---|---|---|---|
| Strongly disagree | 1 | 0 | 0 | 0 | 0 | 0 |
| Disagree | 2 | 2 | 2 | 2 | 2 | 2 |
| Neither | 3 | 2 | 2 | 6 | 1 | 1 |
| Agree | 4 | 13 | 14 | 10 | 14 | 7 |
| Strongly agree | 5 | 8 | 7 | 7 | 8 | 15 |
| Total | | 102 | 101 | 97 | 103 | 110 |
| Mean score | | 4.08 | 4.04 | 3.88 | 4.12 | 4.4 |
| Overall mean score | 4.104 | | | | | |
| Attitude | Positive | Positive | Positive | Positive | Positive | Positive |

**Table AII.2**. Data analysis results of criteria II

| Option | Scale | II.1 | II.2 | II.3 | II.4 | II.5 | II.6 |
|---|---|---|---|---|---|---|---|
| Strongly disagree | 1 | 0 | 0 | 0 | 0 | 0 | 0 |
| Disagree | 2 | 2 | 2 | 2 | 2 | 2 | 2 |
| Neither | 3 | 1 | 1 | 1 | 1 | 1 | 2 |
| Agree | 4 | 11 | 12 | 11 | 9 | 9 | 11 |
| Strongly agree | 5 | 11 | 10 | 11 | 13 | 13 | 10 |
| Total | | 106 | 105 | 106 | 108 | 108 | 104 |
| Mean score | | 4.2 | 4.2 | 4.2 | 4.3 | 4.3 | 4.2 |
| Overall mean score | 4.2 | | | | | | |
| Attitude | Positive | Positive | Positive | Positive | Positive | Positive | Positive |

**Table AII.3.** Data analysis results of criteria III

| Option | Scale | III.1 | III.2 | III.3 | III.4 |
|---|---|---|---|---|---|
| Strongly disagree | 1 | 0 | 0 | 0 | 0 |
| Disagree | 2 | 1 | 2 | 1 | 1 |
| Neither | 3 | 2 | 3 | 3 | 2 |
| Agree | 4 | 10 | 12 | 14 | 8 |
| Strongly agree | 5 | 12 | 8 | 7 | 14 |
| Total | | 108 | 101 | 102 | 110 |
| Mean score | | 4.3 | 4.0 | 4.1 | 4.4 |
| Overall mean score | 4.2 | | | | |
| Attitude | Positive | Positive | Positive | Positive | Positive |

**Table AII.4**. Data analysis results of framework applicability assessment

| Option | Level | IV.1 | IV.2 | IV.3 | IV.4 |
|---|---|---|---|---|---|
| Strongly disagree | 1 | 0 | 0 | 0 | 0 |
| Disagree | 2 | 1 | 3 | 2 | 3 |
| Neither | 3 | 7 | 5 | 6 | 5 |
| Agree | 4 | 14 | 14 | 14 | 14 |
| Strongly agree | 5 | 3 | 3 | 3 | 3 |
| Total | | 94 | 92 | 93 | 92 |
| Mean score | | 3.8 | 3.7 | 3.7 | 3.7 |
| Overall mean score | 3.7 | | | | |
| Attitude | Positive | Positive | Positive | Positive | Positive |

**Table AII.5.** Data analysis results of framework selection intention

| Option | Scale | V.1 | V.2 | V.3 |
|---|---|---|---|---|
| Strongly disagree | 1 | 0 | 0 | 0 |
| Disagree | 2 | 2 | 2 | 2 |
| Neither | 3 | 8 | 6 | 4 |
| Agree | 4 | 12 | 14 | 18 |
| Strongly agree | 5 | 3 | 3 | 1 |
| Total | | 91 | 93 | 93 |
| Mean score | | 3.6 | 3.7 | 3.7 |
| Overall mean score | 3.7 | | | |
| Attitude | Positive | Positive | Positive | Positive |